\documentclass[%
 reprint,
superscriptaddress,
 amsmath,amssymb,
 aps,
]{revtex4-2}

\usepackage{graphicx}
\usepackage{dcolumn}
\usepackage{bm}
\usepackage{braket}
\usepackage{hyperref}
\usepackage{comment}
\usepackage{lipsum}
\usepackage{mathtools}
\usepackage[usenames,dvipsnames]{color}

\newcommand{\ee}{\mathrm{e}}
\newcommand{\ii}{\mathrm{i}}

\def \FigureRabione
{
\begin{figure}[h]
    \centering
    \includegraphics[width=1\linewidth]{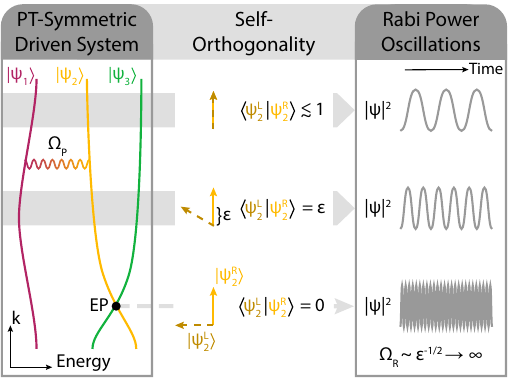}
    \caption{
    Schematic of the consequences of the self-orthogonality on the Rabi frequency $\Omega_\mathrm{R}$. In the Non-Hermitian system, an exceptional point (EP) is formed at the degeneracy of the green and yellow bands and resonant pumping is induced between the violet and yellow bands with pumping frequency $\Omega_\mathrm{P}$. In this setup, the slow power oscillations are dominated by the Rabi frequency $\Omega_\mathrm{R}$. Far away from the EP (top row), the eigenstates $\ket{\psi_i}$ will be less influenced by the EP and the left and right eigenstate of the second band ( $\ket{\psi_2^\mathrm{L}}$ and $\ket{\psi_2^\mathrm{R}}$, respectively) are almost or exactly parallel resulting in a finite $\Omega_\mathrm{R}$. Closer to the EP (middle row), the left and right eigenstate will have smaller overlap $\varepsilon<1$ leading to an increased Rabi frequency. Approaching the EP further, $\varepsilon$ decreases and eventually vanishes exactly at the EP where the eigenstates are self-orthogonal and $\Omega_\mathrm{R}$ diverges.}
    \label{fig_1}
\end{figure}
}
\def \FigureRabitwo
{
\begin{figure}[b!]
    \centering
    \includegraphics[width=1\linewidth]{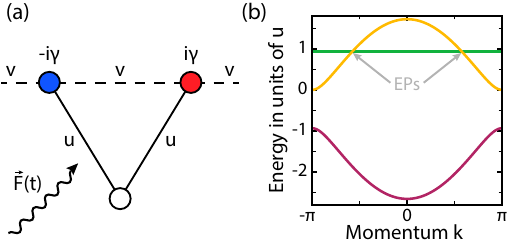}
    \caption{Quasi-one-dimensional three-band model for the realization of diverging Rabi oscillations. (a) Unit cell of the quasi-one-dimensional sawtooth chain consisting of three different sites per unit cell. Red sites are associated with gain ($\ii\gamma$) and blue sites with loss ($-\ii\gamma$). Neutral sites are marked white. Sites with gain and loss are coupled to each other with the hopping strength $v$, and $u$ denotes the hopping amplitude for couplings with the neutral sites. An external force $\vec{F}(t)$ drives the lattice periodically. (b) Energy spectrum of the static $\mathcal{PT}$-symmetric three-band model. The green band is flat and forms the EP with the yellow band at momenta $k^\pm_\mathrm{EP}$.}
    \label{fig_2}
\end{figure}
}

\def \FigureRabithree
{
\begin{figure*}[ht]
    \centering
    \includegraphics[width=1\linewidth]{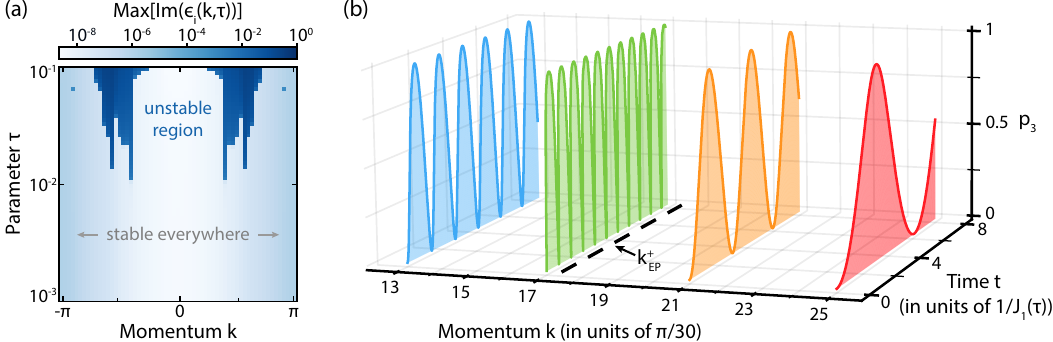}
    \caption{Analysis of the dynamical stability with respect to the driving parameter $\tau$ and diverging Rabi frequency for the bi-orthogonal probability $p_1$. (a) The plot shows the maximal imaginary part of the quasi-energy spectrum $\epsilon_i$ for different values of $\tau$ over the whole BZ. When the imaginary part of the quasi-energies is significant, the time evolution becomes unstable (dark blue area). In the light blue region, the time evolution is stable as the negligible imaginary parts result only from numerical inaccuracies. (b) Time evolution of the bi-orthogonal probability $p_\mathrm{t}$ for four different $k$-values. $p_\mathrm{t}$ describes the probability of detecting the time-evolved state $\ket{\psi(t)}$ in the target eigenstate $\ket{\psi_\mathrm{t}}$. The simulations were performed for a finite chain with PBC consisting of $N=60$ unit cells and the driving parameter $\tau=10^{-3}$. The position of the EP at $k^+_\mathrm{EP}$ is marked by a dashed line. As expected, the oscillation frequency increases with decreasing distance from the EP.}
    \label{fig_3}
\end{figure*}
}

\def \FigureRabifour
{
\begin{figure}
    \centering
    \includegraphics[width=1\linewidth]{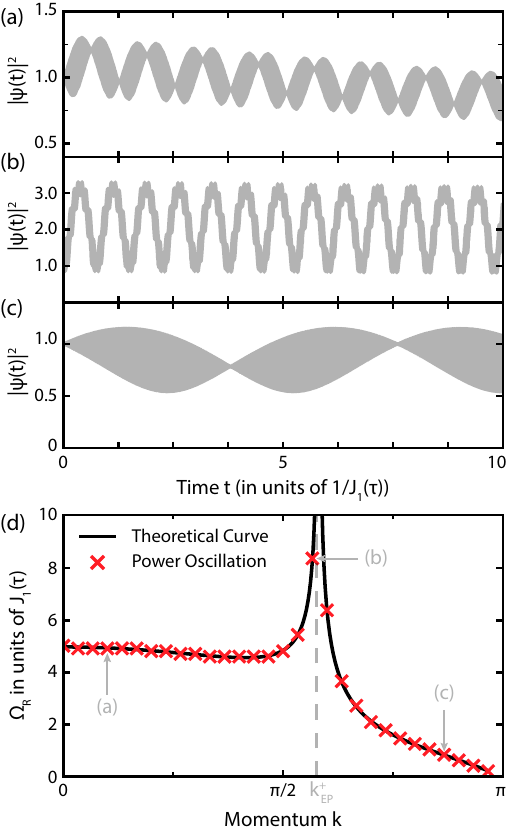}
    \caption{Power oscillations as a function of propagation time. (a), (b) and (c) show the evolution of the total power $\lvert\psi(t)\rvert^2$ as a function of time for three different $k$-values. The fast oscillations are governed by the pumping frequency while the slow oscillations are dominated by the Rabi frequency. (d) Comparison between the analytical functional dependency of the Rabi frequency on the momentum and the numerically determined long time power oscillation frequencies for a finite chain of $N=60$ unit cells and periodic boundary conditions. The theoretically expected Rabi frequency and the power oscillation frequency match very well suggesting that the power oscillations can serve as an observable to unveil the effects of the self-orthogonality at the EP.}
    \label{fig_4}
\end{figure}
}

\begin{document}


\title{Unveiling the Self-Orthogonality at Exceptional Points in Driven $\mathcal{PT}$-Symmetric Systems}
\author{Alexander Fritzsche}
\affiliation{Institute for Theoretical Physics and Astrophysics, University of W\"urzburg, Am Hubland, D-97074 W\"urzburg, Germany}
\affiliation{Institute of Physics, University of Rostock, Rostock, Germany}
\author{Riccardo Sorbello}%
\email{riccardo.sorbello@uni-wuerzburg.de}
\affiliation{Institute for Theoretical Physics and Astrophysics, University of W\"urzburg, Am Hubland, D-97074 W\"urzburg, Germany}
\author{Ronny Thomale}
\affiliation{Institute for Theoretical Physics and Astrophysics, University of W\"urzburg, Am Hubland, D-97074 W\"urzburg, Germany}
\author{Alexander Szameit}
\email{alexander.szameit@uni-rostock.de}
\affiliation{Institute of Physics, University of Rostock, Rostock, Germany}
\affiliation{W\"urzburg-Dresden Cluster of Excellence ct.qmat, D-97074 W\"urzburg, Germany}

\date{\today}

\begin{abstract}
We explore the effect of self-orthogonality at exceptional points (EPs) in non-Hermitian Parity-Time-symmetric systems. Using a driven three-band lattice model, we show that the Rabi frequency diverges as the system approaches an EP due to the coalescence of eigenstates. We demonstrate that this divergence manifests in experimentally accessible power oscillations, establishing a direct observable for self-orthogonality. Our results provide a pathway for probing EP physics in various metamaterial platforms.
\end{abstract}

\maketitle


\section{Introduction}
The description of systems that interact with their environment through non-Hermitian Hamiltonians opened entire research fields in, among others, quantum physics, condensed matter physics or optics~\cite{moiseyev2011non,Ashida02072020,bergholtz2021exceptional}. A characteristic feature of such Hamiltonians is their complex energy spectrum, which, for example, enriches the underlying topology~\cite{Kawabata2019,Cao2021,Wang2021} and leads to new physical phenomena without Hermitian counterpart, like the non-Hermitian skin effect~\cite{CHLee2019,Hofmann2020,Weidemann2020}. However, complex energies also lead to an unstable time evolution as states are exponentially damped or amplified. Hence, the class of open systems with Parity-Time ($\mathcal{PT}$) symmetry was of great interest over the past decades~\cite{BenderPT1998,Bender2015}, as they allow for a real spectrum and therefore, for a stable time evolution. Nevertheless, $\mathcal{PT}$-symmetric systems can still possess characteristic non-Hermitian properties such as exceptional points (EPs) which form when the symmetry breaks.
In parameter space, at an EP not only eigenvalues of the Hamiltonian coincide but also the corresponding eigenstates coalesce. This peculiarity leads to non-trivial winding upon encircling an EP~\cite{Doppler2016,Hassan2017} as well as the phenomenon of self-orthogonality which describes the vanishing of the scalar product of a right eigenstate with its left counterpart~\cite{Sokolov2006,moiseyev2011non,Heiss2012,Lee2020}.
As $\mathcal{PT}$ symmetry and EPs are not restricted to Hamiltonian operators, these concepts have been realized in various experimental platforms~\cite{Chitsazi2017,Feng2017,El-Ganainy2018,Biesenthal2019,Sakhardi2019,Renault2019,Ozdemir2019,Miri2019,Parto2021,Stegmaier2021,Fritzsche2024,Klauck2025} and the unique properties of EPs were exploited to enhance sensing~\cite{Liu2016,Li2016,Wiersig2020,Farhat2020,Rosa2021} and observe unidirectional propagation of light~\cite{Goldzak2018,Koutserimpas2018} and sound~\cite{Shen2018}.
\FigureRabione

In this Letter, we study the effects of the self-orthogonality at EPs on Rabi oscillations in a $\mathcal{PT}$-symmetric driven lattice model. Rabi oscillations have traditionally been studied in the context of quantum control to coherently induce spin flips in two-level systems~\cite{Rabi1937,Braak2016}. In a generalized context, the interplay between Rabi oscillations and $\mathcal{PT}$ symmetry in two-level systems has been extensively examined in the past~\cite{Dietz2007,Joglekar2014,Lee2015,Xie2018,Lu2023,Liu2024,Baradaran2025}. 
Here we employ a different approach based on~\cite{Alfassi2011} where the authors predict the divergence of the Rabi frequency in the vicinity of an EP in optical subwavelength systems based on the formulation of Maxwell's equations as a non-Hermitian operator. We extend this approach to $\mathcal{PT}$-symmetric lattice models where the symmetry enables the emergence of EPs and crucially ensures a stable time evolution.
By showing that the oscillation of the total power, which is another characteristic of $\mathcal{PT}$-symmetric systems, is governed by the Rabi frequency, we propose it as an observable to directly probe the self-orthogonality of eigenstates at the EP. Due to the generality of our results, the model proposed in this Letter can be realized in different metamaterial platforms such as photonic systems or topolectrical circuits.

\section{Rabi Oscillations around an Exceptional Point}
Contrary to Hermitian systems, for a non-Hermitian Hamiltonian $H$ the right eigenvectors defined by $H\ket{\psi^\mathrm{R}_i}=\lambda_i\ket{\psi^\mathrm{R}_i}$ and the left eigenvectors given by $\bra{\psi^\mathrm{L}_i}H=\bra{\psi^\mathrm{L}_i}\lambda_i$ do not coincide. However, if normalized independently, these eigenvectors fulfill the bi-orthogonal condition $\braket{\psi_i^\mathrm{L}\vert\psi_j^\mathrm{R}}=0$ for $i\neq j$ and $\braket{\psi_i^\mathrm{L}\vert\psi_i^\mathrm{R}}=\varepsilon$~\cite{moiseyev2011non,Brody2014,Edvardsson_2023}, allowing for a complete description of non-Hermitian systems.
At an EP ($\lambda_i=\lambda_j$), the Hamiltonian becomes defective and cannot be diagonalized anymore as at least two eigenstates coalesce. As a consequence, $\ket{\psi_j^\mathrm{R}}$ is not only orthogonal to $\bra{\psi^\mathrm{L}_i}$ but also to its own partner $\bra{\psi^\mathrm{L}_j}$ with $\varepsilon\rightarrow 0$. This phenomenon is called self-orthogonality. To create a bi-orthonormal basis with $\braket{\psi_i^\mathrm{L}|\psi_j^\mathrm{R}}=\delta_{ij}$, we choose to renormalize only the left eigenstates $\ket{\psi_i^\mathrm{L}}\to\varepsilon^{-1}\ket{\psi_i^\mathrm{L}}$ following ref.~\cite{Edvardsson_2023}. This normalization choice simplifies some expressions such as the wave function while observables are entirely independent of it (see Supplemental Material~\cite{SM} for details). 
The closer the state is to the EP, the smaller $\varepsilon$ gets, thus resulting in a divergence of the components of the left eigenstate while the bi-orthogonal norm is preserved. To observe this direct consequence of the self-orthogonality, it was proposed to study Rabi oscillations in a subwavelength photonic lattice~\cite{Alfassi2011}. There, the structure of Maxwell's equations results in a non-Hermitian matrix and at the EP forward and backward propagating modes coalesce leading to a divergence of the electric field amplitude and Rabi frequency. 
However, to transfer this electromagnetic phenomenon to lattice systems, we have to realize a dynamically stable three-band model where two bands cross to form a pair of EPs and the third band is only used for the Rabi pumping. Choosing a $\mathcal{PT}$-symmetric system permits EPs to emerge at symmetry breaking points while maintaining a real spectrum. This is necessary to ensure a stable time evolution but not sufficient as external pumping can destabilize the dynamics (see Section~\ref{SecStability} and \cite{SM}).
In the driven, non-Hermitian three-band system, the Rabi frequency $\Omega_\mathrm{R}$ as a function of the Bloch momentum $k$ is given by
\begin{equation}
    \Omega_\mathrm{R}(k)=\sqrt{\left|\braket{\psi  _{j;k}^\mathrm{L}|V(k)|\psi_{i;k}^\mathrm{R}}\right|\cdot\left|\braket{\psi_{i;k}^\mathrm{L}|V(k)|\psi_{j;k}^\mathrm{R}}\right|}\label{eq:rabifrequency}
\end{equation}
where $\ket{\psi_{i;k}^\mathrm{L/R}}$ is the Bloch eigenstate corresponding to the $i$-th band and $V(k)$ describes the interaction due to the pumping between bands $i$ and $j$ (for the full derivation see \cite{SM}). As we induce Rabi oscillations near the EP and only one of the states is self-orthogonal, $\Omega_\mathrm{R}\propto \varepsilon^{-1/2}$ such that it diverges. We show in Section~\ref{power_section} that the total power of the $\mathcal{PT}$-symmetric system oscillates with the same frequency; their relation to self-orthogonality is illustrated schematically in Fig.~\ref{fig_1}.

\section{Diverging Rabi Frequency in a Driven $\mathcal{PT}$-Symmetric Three-Band Model}
\label{SecStability}
\FigureRabitwo
The $\mathcal{PT}$-symmetric three-band model employed to study the described divergence of the Rabi frequency is shown schematically in Fig.~\ref{fig_2}(a). Considering a small external periodic force $\vec{F}(t)$ to induce the pumping, the system is readily described by the effective Bloch Hamiltonian
\FigureRabithree 
\begin{widetext}
\begin{equation}
\begin{split}
    h(t) &=-
    \begin{pmatrix}
        -i\gamma & J_0(\tau)u & J_0(\frac{\tau}{2}
        )v(1+\ee^{-\ii k}) \\
        J_0(\tau)u & 0 & J_0(\tau)u \\
        J_0(\frac{\tau}{2})v(1+\ee^{\ii k}) & J_0(\tau)u & i\gamma
    \end{pmatrix}
    +2\ii\sin\Omega_\mathrm{P}t
    \begin{pmatrix}
        0 & -J_1(\tau)u & - J_1(\frac{\tau}{2})v(1+\ee^{-\ii k}) \\
        J_1(\tau)u & 0 & -J_1(\tau)u \\
        J_1(\frac{\tau}{2})v(1 + \ee^{\ii k}) & J_1(\tau)u & 0
    \end{pmatrix}\\
    &=h_0(k)+\sin\Omega_\mathrm{P}t\cdot V(k)
\end{split}
\label{Hamiltonian}
\end{equation}
\end{widetext}
with the coupling constants $u$ and $v$ and the gain/loss parameter $\gamma=J_0(\tau/2)u\sqrt{2-\frac{J_0(\tau/2)^2u^2}{J_0(\tau)^2v^2}}$. $J_\alpha(\tau)$ are the Bessel functions of the first kind of order $\alpha$ as a function of the perturbation parameter $\tau$. The full derivation of the Hamiltonian, which also includes the Jacobi-Anger expansion in orders of $\tau\ll 1$~\cite{jotzu2014experimental}, can be found in~\cite{SM}. For $\tau=0$, it reduces to the static model discussed in~\cite{Ramezani2017}. The band structure of $h_0(k)$ is shown in Fig.~\ref{fig_2}(b). The spectrum for the tuned value of $\gamma$ is characterized by a flat band (green) and a pair of exceptional points located symmetrically at the momenta
\begin{equation}
    k_\mathrm{EP}^\pm=\pm \arccos \left(\frac{(J_0(\tau/2)u)^4}{(J_0(\tau)v)^4}-1\right).
\end{equation}
In the following, we will focus only on the EP at positive momentum $k_\mathrm{EP}^+$ as the results can be extended symmetrically to its partner $k_\mathrm{EP}^-$. One of the three bands (purple) is separated from the other two by a band gap everywhere in the Brillouin zone (BZ) and serves as the non-self-orthogonal pumping partner.\\
A first study of the Rabi oscillations in this model can be performed by investigating the dynamical evolution of the eigenstates. For that, we consider a finite chain of $N$ unit cells with periodic boundary conditions (PBC) that leads to a discretized momentum space with $k_n=\pi/n$ and integer $n\in \left\lbrace -N/2,...,N/2-1\right\rbrace$. At each fixed momentum, we choose an eigenstate of $h_0(k)$ as the initial state and pump in resonance with the corresponding target eigenstate such that the transition between the states in the bi-orthogonal formalism is described by  
\begin{equation}
    \ket{\psi(t)}=\cos\left(\frac{\Omega_\mathrm{R}}{2} t\right)\ket{\psi_{\mathrm{in}}^\mathrm{R}}+\alpha\sin\left(\frac{\Omega_\mathrm{R}}{2}  t\right)\mathrm{e}^{-\ii\Omega_\mathrm{P} t}\ket{\psi_{\mathrm{t}}^\mathrm{R} }
    \label{rabi_osci}
\end{equation}
where $\ket{\psi_{\mathrm{in}}^\mathrm{R}}$ is the initial right eigenstate and $\ket{\psi_{\mathrm{t}}^\mathrm{R}}$ is the target right eigenstate. The coefficient $\alpha=\bra{\psi_\mathrm{t}^\mathrm{L}}V\ket{\psi_\mathrm{in}^\mathrm{R}}/\bra{\psi_\mathrm{in}^\mathrm{L}}V\ket{\psi_\mathrm{t}^\mathrm{R}}$ is constant when the momentum and the pumping frequency are fixed and diverges when the target state becomes self-orthogonal. Note that the time evolution via the Hamiltonian $h(t)$ does not mix momenta such that the initial and target states, as well as the Rabi and pumping frequency depend on $k$ only parametrically. We therefore omitted the $k$-dependence.

However, a crucial challenge to observe Rabi oscillations in this $\mathcal{PT}$-symmetric model concerns the stability of the system during the time evolution. Although the eigenvalue spectrum of the static model is entirely real, stable dynamics are not guaranteed. We investigate the stability of our model by studying its quasi-energies $\epsilon_i$ which are calculated from the eigenvalues $\mathrm{exp}(-\ii\epsilon_i T)$ of the Floquet operator $U(T)=\mathbb{T}\mathrm{exp}\left(\int_0^T-\ii H(t)\mathrm{d}t\right)$ with $T=2\pi/\Omega_\mathrm{P}$. Here, $\mathbb{T}$ denotes the time-ordering operator and $H(t)$ is the real space Hamiltonian. A stable time evolution is ensured when the imaginary part of $\epsilon_i$ is negligible everywhere in the BZ. Fig.~\ref{fig_3}(a) shows the largest imaginary part of the quasi-energies $\mathrm{Max}[\mathrm{Im(\epsilon)}]$ as a function of momentum $k$ and perturbation parameter $\tau$. We fix the coupling ratio $\frac{v}{u}=1.075$ such that $k_\mathrm{EP}^\pm\neq k_n\, \forall n$ and pump in resonance with the gapped (purple) and highest energy band (either yellow or green depending on $k$), as this further enhances stability (see~\cite{SM}). In the dark blue regions, the time evolution is unstable, while the small imaginary part in the light blue region only results from numerical inaccuracies. Consequently, we choose $\tau=10^{-3}$ in all subsequent simulations.\\
In this stable regime, we perform real space simulations to observe the oscillating behavior of the eigenstates described in Eq.~\eqref{rabi_osci}. As the initial state, we choose the Fourier-transformed gapped right eigenvector of the $h_0(k)$. The time-propagated wave function $\ket{\psi(t)}$ is then Fourier-transformed back and projected onto the unperturbed eigenstates of $h_0(k)$ to calculate the normalized bi-orthogonal transition probabilities $p_i(t) = |c_i(t)|^2/\sum_i|c_i(t)|^2$, where the complex coefficients are defined by $\ket{\psi(t)}=\sum_i c_i(t)\ket{\psi_i^\mathrm{R}}$.
Here, $p_i$ describes the probability of finding $\ket{\psi(t)}$ in the $i$-th eigenstate. The results of the simulations are shown in Fig.~\ref{fig_3}(b) where we show the bi-orthogonal probability of the wave function $\ket{\psi(t)}$ being detected in the target state, $p_\mathrm{t}(t)$, for four different momenta around the EP during the time evolution. We observe very fast oscillations close to the EP (marked with a dashed line) as well as a decreased Rabi frequency farther away from it which shows the diverging characteristics of the Rabi frequency close to the degeneracy.
We also find that close to the EP the pumping efficiency is robust against detuning of the pumping frequency. This is a consequence of the divergent Rabi frequency that can overshadow a higher amount of detuning compared to the Hermitian counterpart (see~\cite{SM}).
\section{Power Oscillations around the Exceptional Point}
\label{power_section}
\FigureRabifour
The divergence of the Rabi frequency discussed in the previous section can only be observed experimentally by resolving the  state spatially for every time step as its Fourier transform is projected onto the Bloch eigenstates to calculate the bi-orthogonal probability $p$. However, as this is challenging in many experimental platforms, we propose the total power as an observable to see the divergence of the Rabi frequency and, by extension, observe the self-orthogonality.

As a consequence of the non-Hermiticity, the time evolution is non-unitary and therefore the total power of the system defined by the square of $L_2$-norm of the wave function $\lVert \ket{\psi(t)}\rVert^2\coloneqq |\psi(t)|^2$ is not preserved. For $\mathcal{PT}$-symmetric systems, however, the real spectrum imposes a quasi-unitary time evolution such that the total power oscillates around an average value~\cite{Bian2020}.
In a setup with Rabi pumping (following from Eq.~\eqref{rabi_osci}), this power oscillation is given by
\begin{equation}
\begin{split}
    |\psi(t)|^2=&\cos^2\left(\frac{\Omega_\mathrm{R}}{2}t\right)|\psi_\mathrm{in}^\mathrm{R}|^2+|\alpha|^2\sin^2\left(\frac{\Omega_\mathrm{R}}{2} t\right) |\psi_\mathrm{t}^\mathrm{R}|^2
\\    &+\sin \left(\Omega_\mathrm{R} t\right)\mathrm{Re}\left[\alpha \ee^{\ii\Omega_\mathrm{P}t}\braket{\psi_\mathrm{t}^\mathrm{R}|\psi_\mathrm{in}^\mathrm{R}}\right].
    \end{split}
    \label{eq.Rabi_norm}
\end{equation}
In an Hermitian system, the last term vanishes as eigenstates are orthogonal and the initial and target states are normalized such that the total power does not oscillate over time. Here, however, there are two independent oscillations shaping the time evolution of the power.
The fast oscillations, on the one hand, are governed by the pumping frequency $\Omega_\mathrm{P}$ and their amplitude is overshadowed close to the EP where the third term in Eq.~\eqref{eq.Rabi_norm} is significantly smaller than the second term considering that $\alpha$ diverges when the target state becomes self-orthogonal. The long-time evolution, on the other hand, is determined by the Rabi frequency $\Omega_\mathrm{R}$, which means that it reflects the influence of the self-orthogonality on $\Omega_\mathrm{R}$. In Fig.~\ref{fig_4}(a)-(c), we show the results for the time evolution of the total power for three exemplary $k$-momenta around the EP at $k_\mathrm{EP}^+$ (marked by the dashed gray line). The initial state is prepared analogously to the previous section, again for a system with PBC and $N=60$ unit cells. Fig.~\ref{fig_4}(d) shows the Rabi frequency as a function of momentum $k$. We observe that the Rabi frequencies obtained from the power oscillations determined via spectral analysis (red crosses) quantitatively match the expected curve obtained from applying Eq.~\eqref{eq:rabifrequency} to our model given in Eq.~\eqref{Hamiltonian} (black line).
These results show that the experimentally more easily accessible total power of the system serves as a reliable global observable for measuring the effects of self-orthogonality at the EP. As the lattice possesses PBC and the states are fully localized in momentum space and, hence, delocalized in real space, the system is fully translationally invariant. Therefore, the power oscillations inside a unit cell follow the same behavior as the total power such that they can be employed as a local observable for the Rabi frequency.

\section{Conclusions}
In this Letter, we showed that the system dynamics can unveil the self-orthogonality phenomenon at exceptional points in setups with Parity-Time symmetry. In the driven three-band model presented in this work, we induced pumping between a gapped band and one of the bands participating in the formation of the EP and showed by tracking the time evolution of the corresponding eigenstates that the resulting Rabi frequency diverges at the degeneracy as long as the dynamics are stable everywhere in the Brillouin zone. As these eigenstate dynamics are difficult to measure in many experimental platforms, we propose the time evolution of the total power as a more useful observable to study the direct consequences of the self-orthogonality. Through analytical and numerical calculations, we showed that the slow oscillations of the power of the $\mathcal{PT}$-symmetric model are governed by the Rabi frequency and therefore show the same diverging behavior in the vicinity of an EP. 
We believe that our work opens the pathway to further exploit the unique properties of EPs that have no Hermitian counterpart. 
\begin{acknowledgments}
The work is funded by the Deutsche Forschungsgemeinschaft (DFG, German Research Foundation) through Project-ID 258499086 - SFB 1170, the W\"urzburg-Dresden Cluster of Excellence on Complexity and Topology in Quantum Matter -- \textit{ct.qmat} Project-ID 390858490 - EXC 2147, Project-ID 441234705 - SFB 1477 ‘Light–Matter Interactions at Interfaces’ and Project-ID 437567992 - IRTG 2676/1-2023 ‘Imaging of Quantum Systems’. A.S. acknowledges funding from the DFG (grants SZ 276/9-2, SZ 276/19-1, SZ 276/20-1, SZ 276/21-1 and SZ 276/27-1), the FET Open Grant EPIQUS (grant no. 899368) within the framework of the European H2020 programme for Excellent Science, as well as the Krupp von Bohlen and Halbach Foundation.
\end{acknowledgments}
\section*{Notes}
A.F. and R.S. contributed equally to this Letter.
\bibliographystyle{apsrev4-2}
\bibliography{bibliography_arxiv.bib}

\appendix

\onecolumngrid

\appendix
\section{Derivation of the Rabi Frequency in a non-Hermitian Model and Power Oscillations}
\subsection{Rabi Oscillations in the Vicinity of an Exceptional Point}

In this section, we want to derive the Rabi frequency in a non-Hermitian system and argue why it diverges in the vicinity of an EP. An important starting remark is that to make the physical observables invariant on the normalization choice of the bi-orthonormal basis, the derivation of physical quantities has to include the metric tensor $G$~\cite{Edvardsson_2023}. The tensor is defined as
\begin{equation}
    G = \sum_n g_n \ket{\psi_n^\mathrm{L}}\bra{\psi_n^\mathrm{L}},
\end{equation}
where the coefficients $g_n$ can be chosen in any way that gives a consistent definition of probability. A possible expression of the coefficients is $g_n=\bra{\psi_n^\mathrm{R}}\psi_n^\mathrm{R}\rangle/|\bra{\psi_n^\mathrm{L}}\psi_n^\mathrm{R}\rangle|^2$. We choose the coefficients that make the metric tensor $G$ diagonal in the basis of the left eigenvectors, \textit{i.e.} $\bra{\psi_n^\mathrm{L}}\psi_n^\mathrm{R}\rangle=1$ and $\bra{\psi_n^\mathrm{R}}\psi_n^\mathrm{R}\rangle=1$, which means that only the left eigenstates carry the signature of self-orthogonality while the right eigenstates have unitary norm. Using this convention, the wave function can simply be written as $\ket{\psi} =\sum_ic_i\ket{\psi_i^\mathrm{R}}$, with $c_i=\bra{\psi_i^\mathrm{R}}G\ket{\psi}=\bra{\psi_i^\mathrm{L}}\psi\rangle$. Similarly, observables are defined as the expectation values of the operator using the metric tensor $G$ as 
\begin{equation}
    \langle \hat{O}\rangle = \frac{\bra{\psi}G\hat{O}\ket{\psi}}{\bra{\psi}G\ket{\psi}}.
    \label{Eq.ExpValue}
\end{equation}
Bearing this in mind, we start with a Bloch Hamiltonian that consists of a static part and a small, time-dependent and translational invariant potential
\begin{align}
   h(k)=h_0(k)+V(k,t).
\end{align}
The unperturbed Hamiltonian then satisfies the eigenvalue relation
\begin{align}
h_0(k)\ket{\psi^\mathrm{R}_{n;k}}=\epsilon_{n;k}\ket{\psi^\mathrm{R}_{n;k}}
\end{align}
with the energy eigenvalues $\epsilon_{n;k}$ and right eigenvectors $\ket{\psi^\mathrm{R}_{n;k}}$. The general time dependent state at a fixed momentum $k$ and under the action of the potential can be decomposed as
\begin{align}
    \ket{\psi_k(t)}=\sum_{n}c_{n;k}(t)\ket{\psi^\mathrm{R}_{n;k}}\ee^{-\ii \epsilon_{n;k}t},
\end{align}
where the time-dependence is accounted for through the phase term and the complex coefficients $c_{n;k}(t)$. Applying this, the Schrödinger equation then becomes
\begin{align}
    \sum_{n}\ii\dot{c}_{n;k}(t)\ket{\psi^\mathrm{R}_{n;k}}\ee^{-\ii \epsilon_{n;k}t}=\sum_{n}V(k,t) c_{n;k}(t)\ket{\psi^\mathrm{R}_{n;k}}\ee^{-\ii \epsilon_{n;k}t}.
\end{align}
After multiplying the Schrödinger equation with the left eigenvector $\ee^{\ii\epsilon_{n';k}t}\bra{\psi_{n';k}^\mathrm{L}}$ and using the bi-orthogonal relation $\braket{\psi^\mathrm{L}_{n';k}|\psi^\mathrm{R}_{n;k}}=\delta_{n,n'}$ it becomes
\begin{align}
\ii\dot{c}_{n';k}=\sum_{n}\braket{\psi^\mathrm{L}_{n';k}|V(k,t)|\psi^\mathrm{R}_{n;k}}\ee^{-\ii(\epsilon_{n;k}-\epsilon_{n';k})t}.
\end{align}
We now write $\epsilon_{n;k}-\epsilon_{n';k}=\Delta\epsilon$ and use separation of variables $V(k,t)=V(k)\Theta(t)$ with the time-dependent term $\Theta(t)=\sin(\Omega_\mathrm{P}t)=\frac{1}{2\ii}(\ee^{\ii\Omega_\mathrm{P}t}-\ee^{-\ii\Omega_\mathrm{P}t})$ being periodic with the pumping frequency $\Omega_\mathrm{P}$ to obtain
\begin{align}
    \ii\dot{c}_{n';k}=\frac{1}{2\ii}\sum_{n}\braket{\psi^\mathrm{L}_{n';k}|V(k)|\psi^\mathrm{R}_{n;k}}\left(\ee^{-\ii(\Delta\epsilon-\Omega_\mathrm{P})t}-\ee^{-\ii(\Delta\epsilon+\Omega_\mathrm{P})t}\right).
\end{align}
In the following, we employ resonant pumping such that $\Omega_\mathrm{P}= \Delta\epsilon$ and apply the rotating wave approximation to assume that the oscillation $\propto 2\Omega_\mathrm{P}$ averages out. When we now pump between an initial and target state, we obtain a system of two coupled equations
\begin{align}
    \ii\dot{c}_{\mathrm{in};k}(t)=c_{\mathrm{t};k}(t)\frac{1}{2\ii}\braket{\psi^\mathrm{L}_{\mathrm{in};k}|V(k)|\psi^\mathrm{R}_{\mathrm{t};k}}\\
    \ii\dot{c}_{\mathrm{t};k}(t)=-c_{\mathrm{in};k}(t)\frac{1}{2\ii}\braket{\psi^\mathrm{L}_{\mathrm{t};k}|V(k)|\psi^\mathrm{R}_{\mathrm{in};k}}.
    \label{set_equ_Rabi}
\end{align}
We solve the coupled differential equations with the boundary conditions $c_\mathrm{in;k}(0)=1$ and $c_\mathrm{t;k}(0)=0$ to obtain 
\begin{align}
    c_{\mathrm{in};k}(t) &=  \cos \left(\frac{\Omega_\mathrm{R}}{2}t  \right) \\
    c_{\mathrm{t};k}(t) &=  \frac{\bra{\psi_{\mathrm{t};k}^L}V(k)\ket{\psi_{\mathrm{in};k}^R}}{\bra{\psi_{\mathrm{in};k}^L}V(k)\ket{\psi_{\mathrm{t};k}^R}}\sin \left( \frac{\Omega_\mathrm{R}}{2}t\right) := \alpha \sin \left(\frac{\Omega_\mathrm{R}}{2}t\right).
\end{align}
The Rabi frequency $\Omega_\mathrm{R}$ then reads
\begin{align}
\Omega_\mathrm{R}=\sqrt{|\braket{\psi^\mathrm{L}_{\mathrm{t};k}|V(k)|\psi^\mathrm{R}_{\mathrm{in};k}}|\cdot|\braket{\psi^\mathrm{L}_{\mathrm{in};k}|V(k)|\psi^\mathrm{R}_{\mathrm{t};k}}|}.
\label{Rabi_Frequ_SM}
\end{align}
This frequency diverges if the initial or target state becomes self-orthogonal. As mentioned in the main text, the normalization choice influences the shape of the coefficients (since we write the same state in a different basis) while the Rabi frequency is an observable and remains unchanged.  
\subsection{Oscillations of the Total Intensity in a non-Hermitian system with Rabi pumping}
In this section, we show that the slow oscillation of the total power in the $\mathcal{PT}$-symmetric setup is determined by the Rabi frequency. In a non-Hermitian system where we induce resonant pumping between two bands starting from an initial state $\ket{\psi_{\mathrm{in};k}}$ and pumping to a target state $\ket{\psi_{\mathrm{t};k}}$, the time evolution, analogously to the Hermitian Rabi oscillations, is given by 
\begin{align}
    \ket{\psi_k(t)}=\cos\left(\frac{\Omega_\mathrm{R}}{2} t\right)\ket{\psi_{\mathrm{in};k}^\mathrm{R}}+\alpha \sin\left(\frac{\Omega_\mathrm{R}}{2}  t\right)\mathrm{e}^{-\ii\Omega_\mathrm{P} t}\ket{\psi_{\mathrm{t};k}^\mathrm{R} }
\end{align}
Taking into account that the initial and target states are non-orthogonal in the non-Hermitian system and that the normalization is between left and right eigenstates, we write the power $|\psi_k(t)|^2$ as
\begin{align}
\begin{split}
    |\psi_k(t)|^2=&\cos^2\left(\frac{\Omega_\mathrm{R}}{2} t\right) |\psi^\mathrm{R}_{\mathrm{in};k}|^2+|\alpha|^2\sin^2\left(\frac{\Omega_\mathrm{R}}{2} t\right) |\psi^\mathrm{R}_{\mathrm{t};k}|^2+\alpha\cos\left(\frac{\Omega_\mathrm{R}}{2} t\right)\sin\left(\frac{\Omega_\mathrm{R}}{2} t\right)\mathrm{e}^{\ii\Omega_\mathrm{P}t}\braket{\psi^\mathrm{R}_{\mathrm{t};k}|\psi^\mathrm{R}_{\mathrm{in};k}}\\
    &+\alpha^* \cos\left(\frac{\Omega_\mathrm{R}}{2} t\right)\sin\left(\frac{\Omega_\mathrm{R}}{2} t\right)\mathrm{e}^{-\ii\Omega_\mathrm{P}t}\braket{\psi^\mathrm{R}_{\mathrm{in};k}|\psi^\mathrm{R}_{\mathrm{t};k}}
\end{split}
\end{align}
with $\braket{\psi^\mathrm{R}_{\mathrm{in};k}|\psi^\mathrm{R}_{\mathrm{t};k}}=\braket{\psi^\mathrm{R}_{\mathrm{t};k}|\psi^\mathrm{R}_{\mathrm{in};k}}^*$ the last term becomes the complex conjugate of the second to last term such that we can write
\begin{align}
\begin{split}
    |\psi_k(t)|^2&=\cos^2\left(\frac{\Omega_\mathrm{R}}{2} t\right) |\psi^\mathrm{R}_{\mathrm{in};k}|^2+|\alpha|^2\sin^2\left(\frac{\Omega_\mathrm{R}}{2} t\right) |\psi^\mathrm{R}_{\mathrm{t};k}|^2+2\cos\left(\frac{\Omega_\mathrm{R}}{2} t\right)\sin\left(\frac{\Omega_\mathrm{R}}{2} t\right)\mathrm{Re}\left[\alpha\mathrm{e}^{\ii\Omega_\mathrm{P}t}\braket{\psi^\mathrm{R}_{\mathrm{t};k}|\psi^\mathrm{R}_{\mathrm{in};k}}\right]\\
    &=\cos^2\left(\frac{\Omega_\mathrm{R}}{2} t\right) |\psi^\mathrm{R}_{\mathrm{in};k}|^2+|\alpha|^2\sin^2\left(\frac{\Omega_\mathrm{R}}{2} t\right) |\psi^\mathrm{R}_{\mathrm{t};k}|^2+\sin\left( \Omega_\mathrm{R} t\right)\mathrm{Re}\left[\alpha\mathrm{e}^{\ii\Omega_\mathrm{P}t}\braket{\psi^\mathrm{R}_{\mathrm{t};k}|\psi^\mathrm{R}_{\mathrm{in};k}}\right]
\end{split}
 \label{Rabi_power_SM}
\end{align}
The exponential term contains the fast oscillations of the driving while slow oscillations are determined by the Rabi frequency $\Omega_\mathrm{R}$.

\section{The Periodically Driven $\mathcal{PT}$-Symmetric Three Band Model}
\subsection{Analytical Calculations for the Static Non-Hermitian Three-Band Model}
\begin{figure}
    \centering
    \includegraphics[width=0.5\linewidth]{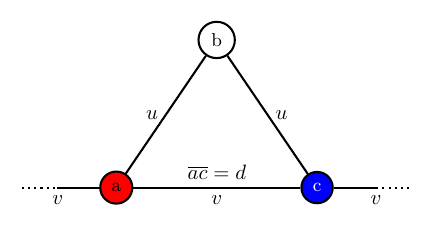}
    \caption{Schematic of the unit cell of the static $\mathcal{PT}$-symmetric three-band model. The red sites ($a$) are associated with gain and the blue sites ($b$) with loss. The white sites ($b$) are neutral. $u$ and $v$ denote the different coupling constants and the distance between sites $a$ and $c$ is denoted by $d$.}
    \label{SM_lattice}
\end{figure}

In this section, we repeat the analytical calculations of the static $\mathcal{PT}$-symmetric three-band model presented in \cite{Ramezani2017}. \\
The unit cell of the model with the different couplings $u$ and $v$ and the three sublattices $a,b,c$ are presented in Fig.\ref{SM_lattice}. Gain ($+\ii\gamma$) is associated with the sites $a$ and loss ($-\ii\gamma$) with sites $c$.
The wavefunction of the $n$-th unit cell $\psi_n=(a_n,b_n,c_n)^\mathrm{T}$ is given through the coupled differential equations 
\begin{equation}
    \begin{split}
	\ii\frac{\mathrm{d}a_n}{\mathrm{d}t}=\ii\gamma a_n-u\,b_n-v(c_n+c_{n-1}) \\
	\ii\frac{\mathrm{d} b_n}{\mathrm{d} t}=-u(a_n+c_n)\\
	\ii\frac{\mathrm{d} c_n}{\mathrm{d} t}=-\ii\gamma c_n-u b_n-v(a_n+a_{n+1})
	\end{split}
\end{equation}
and the Bloch Hamiltonian is given by
\begin{equation}
    h^\mathrm{static}_0(k)=\left(\begin{array}{ccc}
	\ii \gamma & -u & -v(1+\mathrm{e}^{-\ii k}) \\ 
	-u & 0 & -u \\ 
	-v(1+\mathrm{e}^{\ii k}) & -u & -\ii \gamma
	\end{array} \right).
\end{equation}
To ensure a stable time evolution in this model in the presence of a degeneracy, one band needs to be flat, which results in the formation of EPs as well as an entirely real spectrum. We now want to compute the right eigenvector that corresponds to this flat band at $\epsilon_0$ by solving
\begin{equation}
\left(\begin{array}{ccc}
	\ii\gamma-\epsilon_0 & -u & -v(1+\mathrm{e}^{-\ii k}) \\ 
	-u & -\epsilon_0 & -u \\ 
	-v(1+\mathrm{e}^{\ii k}) & -u & -\ii\gamma -\epsilon_0
	\end{array} \right)\left(\begin{array}{c}
	\alpha_1 \\ 
	\alpha_2 \\ 
	\alpha_3
	\end{array} \right)=0
\end{equation}
and obtain the two expressions
\begin{equation}
	\ket{\psi^\mathrm{R}_{\epsilon_0}}=\left(\begin{array}{c}
	1 \\ 
	-\frac{u}{\epsilon_0}(1+\chi^\mathrm{R}) \\ 
	\chi^\mathrm{R}
	\end{array} \right)\quad \mathrm{or} \quad \ket{\psi^{'\mathrm{R}}_{\epsilon_0}}=\left(\begin{array}{c}
	1 \\ 
	-\frac{u}{\epsilon_0}(1+\frac{1}{\chi^{\mathrm{R}*}}) \\ 
	\frac{1}{\chi^{\mathrm{R}*}}
	\end{array} \right)
	\end{equation}
    with $\chi^\mathrm{R}=\frac{-\ii\gamma+\epsilon_0-\frac{u^2}{\epsilon_0}}{\frac{u^2}{\epsilon_0}-v(1+\mathrm{e}^{-\ii k})}$. By assuming that the flat band participates in the formation of the EPs at some $k_\mathrm{EP}$ we need to impose these two expressions have to be equal as despite having a degeneracy there can only be one linearly independent eigenvector. This gives us a condition for the gain/loss parameter.
    It is given by
    \begin{equation}
        \gamma_{\epsilon_0}=\frac{1}{\sqrt{\epsilon_0}}\sqrt{2u^2\epsilon_0+2v^2\epsilon_0-\epsilon_0^3+2\cos(k_\mathrm{EP})v(v\epsilon_0-u^2)-2u^2v}.
    \end{equation}
    
    Using this value for $\gamma$, we employ numerical methods to solve the characteristic polynomial for $\gamma=\gamma_{\epsilon_0}$
    \begin{equation}
        \epsilon^3+\epsilon(\gamma_{\epsilon_0}^2-2u^2-2v^2-2v^2\cos(k_\mathrm{EP}))+2vu^2\cos(k)+2vu^2=0
    \end{equation}
    which results in the three eigenvalues
    \begin{equation}
        \epsilon=\left[\epsilon_0\, ; \, -\frac{\epsilon_0}{2}\pm \sqrt{\frac{\epsilon_0^2}{4}+\frac{4vu^2}{\epsilon_0}\cos^2(k_\mathrm{EP}/2)}\right].
    \end{equation}
The EPs can be found at the degeneracy of $\epsilon_0$ and $-\frac{\epsilon_0}{2}+ \sqrt{\frac{\epsilon_0^2}{4}+\frac{4vu^2}{\epsilon_0}\cos^2(k_\mathrm{EP}/2)}$. They are located at
\begin{equation}
    k_\mathrm{EP}^\pm=\pm \arccos\left(\frac{\epsilon_0^3-vu^2}{vu^2}\right).
\end{equation}
From further numerical analysis, we find that the flat band is located at $\epsilon_0=u^2/v$. Using this and the momenta $k_\mathrm{EP}$ we can calculate the gain/loss parameter $\gamma_{\epsilon_0}$ to find
\begin{equation}
    \gamma_{\epsilon_0}=u\sqrt{2-\frac{u^2}{v^2}}
\end{equation}
which leads to the constraints needed to ensure the formation of EPs $\left|\frac{u^4}{v^4}-1\right|\leq 1$ and $\frac{u^2}{v^2}< 2$. Note that by imposing the coalescence of eigenstates the degeneracy will be an EP.

\subsection{Effective Hamiltonian of the Weakly Driven $\mathcal{PT}$-Symmetric Three-Band System}
To derive the effective Hamiltonian for the weakly periodically driven three-band model, whose static properties were discussed in the previous section, we start from the real space tight-binding Hamiltonian
	\begin{equation}
	\begin{split}
	H(t)=\sum_i (\ii\gamma a^\dagger_i a_i - \ii\gamma c^\dagger_i c_i) + \sum_{i } (-v a^\dagger_ic_i  -u a^\dagger_i b_i -ub^\dagger_i c_i +\mathrm{h.c.} -va^\dagger_ic_{i-1}-vc^\dagger_ia_{i+1})
	+\sum_{i,s\in \left\lbrace a,b,c\right\rbrace} (\vec{F}(t)\cdot \vec{r}_i^s)s^\dagger_i s_i
	\end{split}
	\end{equation}
where $a^{(\dagger)},b^{(\dagger)},c^{(\dagger)}$ denote the annihilation (creation) operators on the corresponding sublattice sites and $\vec{F}(t)$ is a weak periodic force that acts as an on-site term on the sites $s\in \left\lbrace a,b,c\right\rbrace$. $\vec{r}_i^s$ denotes the position of the $i$-th lattice site of type $s=a,b,c$ in the lattice frame. The force is given by $\vec{F}(t)=-m\ddot{\vec{r}}_\mathrm{lat}(t)$ with some mass parameter $m$. The subsequent derivation follows the calculations performed in~\cite{jotzu2014experimental}. We start by performing the unitary transformation
\begin{equation}
\mathcal{U}=\mathrm{exp}\left(\ii\sum_{i,s\in \left\lbrace a,b,c\right\rbrace}(-m\dot{\vec{r}}_\mathrm{lat}\cdot \vec{r}^s_i)s^\dagger_is_i\right)=1+i\sum_{i,s\in \left\lbrace a,b,c\right\rbrace}(-m\dot{\vec{r}}_\mathrm{lat}\cdot \vec{r}^s_i)s^\dagger_is_i-\left[\sum_{i,s\in \left\lbrace a,b,c\right\rbrace}(-m\dot{\vec{r}}_\mathrm{lat}\cdot \vec{r}^s_i)s^\dagger_is_i\right]^2+\dots
\end{equation}
With $\mathcal{U}^\dagger=\mathrm{exp}(-\ii\sum_{i,s\in \left\lbrace a,b,c\right\rbrace}(-m\dot{\vec{r}}_\mathrm{lat}\cdot \vec{r}^s_i)s^\dagger_is_i)$ and $\partial_t\mathcal{U}=i[\sum_{i,s\in \left\lbrace a,b,c\right\rbrace}(\vec{F}(t)\cdot\vec{r}_i^s)s^\dagger_is_i] U$ 
we write the new Hamiltonian as 
\begin{equation}
    H'(t)=\mathcal{U}H(t)\mathcal{U}^\dagger+\ii \partial_t\mathcal{U}\cdot\mathcal{U}^\dagger.
    \label{SM_Ham_Unitarytrafo}
\end{equation}
The last term becomes
\begin{equation}
	\begin{split}
	\partial_t\mathcal{U}\cdot\mathcal{U}^\dagger=\ii\left[\sum_{i,s\in \left\lbrace a,b,c\right\rbrace}(\vec{F}(t)\cdot\vec{r}_i^s)s^\dagger_is_i\right]\cdot\left(1+\ii\sum_{i,s\in \left\lbrace a,b,c\right\rbrace}(-m\dot{\vec{r}}_\mathrm{lat}\cdot \vec{r}^s_i)s^\dagger_is_i+\dots\right)\\\cdot \left(1-\ii\sum_{j,s'\in \left\lbrace a,b,c\right\rbrace}(-m\dot{\vec{r}}_\mathrm{lat}\cdot \vec{r}^{s'}_j)s'^\dagger_js'_j+\dots\right)=\ii\left[\sum_{i,s\in \left\lbrace a,b,c\right\rbrace}(\vec{F}(t)\cdot\vec{r}_i^s)s^\dagger_is_i\right]
	\end{split}
\end{equation}
using $s'_is^\dagger_j=\delta^{ss'}_{ij}$. The first term of Eq.~\eqref{SM_Ham_Unitarytrafo} 
\begin{equation}
	\begin{split}
	\left(1+\ii\sum_{i,s\in \left\lbrace a,b,c\right\rbrace}(-m\dot{\vec{r}}_\mathrm{lat}\cdot \vec{r}^s_i)s^\dagger_is_i+\dots\right)H(t)\left(1-\ii\sum_{j,s'\in \left\lbrace a,b,c\right\rbrace}(-m\dot{\vec{r}}_\mathrm{lat}\cdot \vec{r}^{s'}_j)s'^\dagger_js'_j+\dots\right)
	\end{split}
	\end{equation}
can be simplified using the relations (that can be extended analogously for higher order terms)
\begin{equation}
	\begin{split}
	+\ii\sum_{s',j}(-m\dot{\vec{r}}_\mathrm{lat}\cdot \vec{r}^{s'}_j)s'^\dagger_js'_j \cdot e^\dagger_kf_l+e^\dagger_kf_l\cdot(-\ii\sum_{s,i}(-m\dot{\vec{r}}_\mathrm{lat}\cdot \vec{r}^s_i)s^\dagger_is_i)=\\
	-\ii(-\vec{q}_\mathrm{lat}\cdot\vec{r}^e_k)e^\dagger_kf_l+\ii(-\vec{q}_\mathrm{lat}\cdot\vec{r}^f_l)e^\dagger_kf_l=\ii\vec{q}_\mathrm{lat}\cdot(\vec{r}^e_k-\vec{r}^f_l)e^\dagger_kf_l
	\end{split}
	\end{equation}
where $e^{(\dagger)},f^{(\dagger)}$ denote symbolically the different annihilation and creation operators contained in $H(t)$. Here, we introduced the shorter notation $\vec{q}_\mathrm{lat}=m\dot{\vec{r}}_\mathrm{lat}$. This, consequently, now leads to the new Hamiltonian
\begin{equation}
    \begin{split}
	H'(t)=&\sum_i (\ii\gamma a^\dagger_i a_i - \ii\gamma c^\dagger_i c_i) + \sum_{i } (-v \mathrm{e}^{\ii(\vec{q}_\mathrm{lat}\cdot\vec{r}_{ii}^{ac})} a^\dagger_ic_i  -u\mathrm{e}^{\ii(\vec{q}_\mathrm{lat}\cdot\vec{r}_{ii}^{ab})} a^\dagger_i b_i \\&
	-u\mathrm{e}^{\ii(\vec{q}_\mathrm{lat}\cdot\vec{r}_{ii}^{bc})}b^\dagger_i c_i  -v \mathrm{e}^{\ii(\vec{q}_\mathrm{lat}\cdot\vec{r}_{ii}^{ca})} c^\dagger_ia_i-u\mathrm{e}^{\ii(\vec{q}_\mathrm{lat}\cdot\vec{r}_{ii}^{ba})} b^\dagger_i a_i-u\mathrm{e}^{\ii(\vec{q}_\mathrm{lat}\cdot\vec{r}_{ii}^{cb})}c^\dagger_i b_i \\&-v\mathrm{e}^{\ii(\vec{q}_\mathrm{lat}\cdot\vec{r}_{ii-1}^{ac})}a^\dagger_ic_{i-1}-v\mathrm{e}^{\ii(\vec{q}_\mathrm{lat}\cdot\vec{r}_{ii+1}^{ca})}c^\dagger_ia_{i+1})
	\end{split}
\end{equation}
with $\vec{r}^{ss'}_{ij}=\vec{r}_i^{s}-\vec{r}_j^{s'}$ and $\vec{r}^{ss'}_{ij}=-\vec{r}^{s's}_{ij}$ and $\vec{r}_{ii-1}^{ac}=-\vec{r}_{ii+1}^{ca}$. We hence observe that a modulation of the position of the lattice site corresponds to a modulation of the couplings of the unperturbed lattice. We now assume a periodic modulation of the lattice of the form
\begin{equation}
	\vec{r}_\mathrm{lat}(t)=-A\cos(\Omega_\mathrm{P} t)\vec{e}_1
	\end{equation}
	With such a modulation the exponents become
	\begin{equation}
	\ii(\vec{q}_\mathrm{lat}\cdot\vec{r}_{ij}^{ss'})=\ii mA\Omega_\mathrm{P} (\vec{e}_1\cdot\vec{r}_{ij}^{ss'})\sin(\Omega_\mathrm{P} t)
	\end{equation}
We now define the new variable $w_{ij}^{ss'}=m\Omega_\mathrm{P} A(\vec{e}_1\cdot\vec{r}_{ij}^{ss'})$ for simplicity and therefore write
	\begin{equation}
\mathrm{e}^{\ii(\vec{q}_\mathrm{lat}\cdot\vec{r}_{ij}^{ss'})}=\mathrm{e}^{\ii w_{ij}^{ss'}\sin(\Omega_\mathrm{P} t)}
	\label{couplfull}
	\end{equation}
	We now use the Jacobi-Anger-expansion 
	\begin{equation}
	\mathrm{e}^{\ii t\sin\theta}=\sum_{\alpha=-\infty}^{\infty} J_\alpha(t)\mathrm{e}^{\ii \alpha\theta}
	\end{equation}
	where $J_\alpha(t)$ are the Bessel functions and obtain for the coefficients
    \begin{equation}
	\mathrm{e}^{\ii w_{ij}^{ss'}\sin(\Omega_\mathrm{P} t)}=\sum_{\alpha=-\infty}^{\infty}J_\alpha(w_{ij}^{ss'})\mathrm{e}^{\ii \alpha\Omega_\mathrm{P} t}
	\end{equation} 
    In the following we are only interested in the first two leading orders ($\alpha=\pm 1,0$) as the higher order terms oscillate much faster and scale approximately with $\frac{1}{2^\alpha\alpha!}$. We therefore obtain
    \begin{equation}
	\mathrm{e}^{\ii w_{ij}^{ss'}\sin(\Omega_\mathrm{P} t)}\approx J_0(w_{ij}^{ss'})+J_{-1}(w_{ij}^{ss'})\mathrm{e}^{-\ii\Omega_\mathrm{P} t}+J_1(w_{ij}^{ss'})\mathrm{e}^{\ii\Omega_\mathrm{P} t}=J_0(w_{ij}^{ss'})+2\ii J_{1}(w_{ij}^{ss'})\sin(\Omega_\mathrm{P} t)
	\end{equation}
	where we used the property of the Bessel functions $J_{-\alpha}(t)=(-1)^\alpha J_{\alpha}(t)$. We then finally obtain the modified real space Hamiltonian
    \begin{equation}
	\begin{split}
	H'(t)=&\sum_i (\ii\gamma a^\dagger_i a_i - \ii\gamma c^\dagger_i c_i) +\sum_i [-J_0(w_{ii}^{ac})v(a^\dagger_ic_i+c^\dagger_ia_i)-J_0(w_{ii}^{ab})u(a^\dagger_ib_i+b^\dagger_ia_i)
	\\&-J_0(w_{ii}^{bc})u(b^\dagger_ic_i+c^\dagger_ib_i)-J_0(w_{ii- 1}^{ac})v(a^\dagger_ic_{i-1}+c^\dagger_ia_{i+1})] 
	\\&+\sum_i 2\ii\sin(\Omega_\mathrm{P}t)[-J_1(w_{ii}^{ac})v(a_i^\dagger c_i -c_i^\dagger a_i)-J_1(w_{ii}^{ab})u(a^\dagger_i b_i-b^\dagger_i a_i)-J_1(w_{ii}^{bc})u(b^\dagger_i c_i- c^\dagger_i b_i)
	\\&-J_1(w_{ii- 1}^{ac})v(a^\dagger c_{i-1}-c^\dagger_ia_{i+1})]
	\end{split}
	\end{equation}
    with the relations $J_0(t)=J_0(-t)$ and $J_{1}(t)=-J_1(-t)$. This real space Hamiltonian then translates to the Bloch Hamiltonian given in the main text of our manuscript
    \begin{equation}
	\begin{split}
	h(t)=&\left(\begin{array}{ccc}
	\ii\gamma& -J_0(w_{ii}^{ab})u & -J_0(w_{ii}^{ac})v  -J_0(w_{ii-1}^{ac})v\mathrm{e}^{-\ii k} \\ 
	-J_0(w_{ii}^{ab})u&0  & -J_0(w_{ii}^{bc})u   \\ 
	-J_0(w_{ii}^{ac})v-J_0(w_{ii-1}^{ac})v\mathrm{e}^{\ii k}& -J_0(w_{ii}^{ab})u &-\ii\gamma     
	\end{array} \right) 
	\\&+ 2\sin (\Omega_\mathrm{P} t)\left( \begin{array}{ccc}
	0& -\ii J_1(w_{ii}^{ab})u  & -\ii J_1(w_{ii}^{ac})v-\ii J_1(w_{ii- 1}^{ac})v\mathrm{e}^{-\ii k}  \\ 
	\ii J_1(w_{ii}^{ab})u&  0&-\ii J_1(w_{ii}^{bc})u    \\ 
	 \ii J_1(w_{ii}^{ac})v+\ii J_1(w_{ii- 1}^{ac})v\mathrm{e}^{\ii k} & \ii J_1(w_{ii}^{bc})u &0    
	\end{array}  \right).
	\label{H0appr}
	\end{split}
	\end{equation}
    For simplicity, in the main text, we define the driving or perturbation parameter $\tau$ in this quasi-one-dimensional lattice consisting of isosceles triangles
    \begin{equation}
	\begin{split}
	w_{ii}^{ac}=w_{ii-1}^{ac}=2w_{ii}^{ab}=2w_{ii}^{bc}=mA\Omega_\mathrm{P} d=\tau.	
	\end{split}
	\end{equation}
    We therefore obtain for the modified conditions for the existence of an exceptional point (given in the previous section)
	\begin{equation}
	\begin{split}
	\left|\frac{J_0(\tau/2)^4u^4}{J_0(\tau)^4v^4}-1\right|\leq 1 \quad \mathrm{and}\quad \frac{J_0(\tau/2)^2u^2}{J_0(\tau)^2v^2}<2
	\end{split}
	\end{equation}
	and 
	\begin{equation}
	\gamma=J_0(\tau/2)u\sqrt{2-\frac{J_0(\tau/2)^2u^2}{J_0(\tau)^2v^2}}\quad\mathrm{and}\quad k_\mathrm{EP}=\pm \arccos\left(\frac{J_0(\tau/2)^4u^4}{J_0(\tau)^4v^4}-1\right).
	\end{equation}
\section{Study of the dynamical stability}
\label{App.Stability}

Although $\mathcal{P}\mathcal{T}$ symmetry can bound the system to have only real energies, it does not ensure dynamical stability. In particular, parametric resonances or instantaneous $\mathcal{PT}$ symmetry breaking can cause instabilities when the system is driven. Since the system is periodically driven the stability is determined by the associated Floquet multipliers, \textit{i.e.} the eigenvalues of the Floquet operator
\begin{equation}
    \hat{U}(T) = \mathbb{T} \exp \left( \int_0^T -\ii \hat{H}(t) \mathrm{d}t \right) = \lim_{N\to\infty} \prod_{n=0}^{N-1} \exp \left( -\ii \, \hat{H}\left(\dfrac{nT}{N}\right)\, \dfrac{T}{N} \right) \, ,
    \label{eq.floq_op}
\end{equation}
where $\mathbb{T}$ is the time ordering operator. The system is dynamically stable if the Floquet multipliers are just phases, meaning that the associated dynamics are unitary. When $\mathcal{PT}$ symmetry is, even instantaneously, broken at least two roots of the characteristic polynomial become complex conjugate with a damped mode and an exponentially growing one. We consider the driven Bloch Hamiltonian $h_k(t)$ of Eq.~\eqref{Hamiltonian} and calculate the corresponding Floquet operator by using Eq.~\eqref{eq.floq_op}, thus obtaining the Bloch-Floquet operator $u(T,k)$. If $\lambda_j(k)$ are the Floquet multipliers, \textit{i.e.} the eigenvalues of $u(T,k)$, the corresponding quasi-energies are defined through $\lambda_j(k)=e^{-\ii \epsilon_j(k) T}$. The quasi-energies $\epsilon_j(k)$ are defined in the first Floquet-Bloch zone (FFBZ), as they are invariant modulo-$\frac{2\pi}{T}$ and the Bloch momentum is defined in $[-\pi,\pi[$. Hence, the quasi-energies in the FFBZ wrap around both the $k$ and the $\epsilon(k)$ axes. Furthermore, the quasi-energies $\epsilon(k)$ can be interpreted as the eigenvalues of the Floquet Hamiltonian $H_\mathrm{F}$. The Floquet Hamiltonian is defined as $H_\mathrm{F} := \frac{\ii}{T}\log U(T) $, and gives the same long-scale dynamics of $H(t)$ while being time-independent, as the short-scale dynamics have already been integrated.  \\
\begin{figure}[ht]
    \centering
    \includegraphics[width=\linewidth]{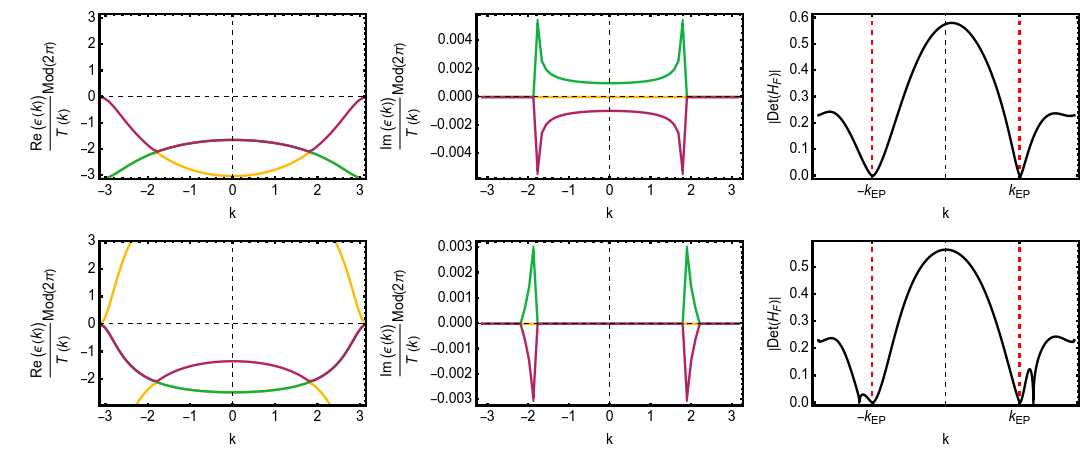}
    \caption{Real and imaginary part of the quasi-energy bands $\epsilon(k)$ and  The top row is calculated when pumping from the first band (purple in Fig.~\ref{fig_2}) to the flat band (green in Fig.~\ref{fig_2}) while the bottom row represents the case of pumping between the first band and the top dispersive band (yellow in Fig.~\ref{fig_2}). In both cases the quasi-energy bands exhibit exceptional points at $k_{EP}$. However, for the flat band we observe a $\mathcal{PT}$-symmetry breaking between the pair of EPs while for the dispersive band the $PT$-broken region lies outside the EPs.}
    \label{fig:quasi-energy-band}
\end{figure}
As extensively studied in~\cite{Longhi_2017}, the Floquet Hamiltonian $H_\mathrm{F}$ can show emergent properties and characteristics that are not shared with the static Hamiltonian but are given by the driving term. In particular, in the starting model the fine-tuned gain and loss terms ensure an entirely real spectrum with $\mathcal{PT}$ symmetry breaking only at the two exceptional points. Here we numerically calculate the Floquet Hamiltonian in the two possible scenarios: when pumping from the gapped band to the flat band or when pumping from the gapped band to the other dispersive band. As we can see in Fig.~\ref{fig:quasi-energy-band} (right row), in both cases the determinant of the Floquet Hamiltonian vanishes at $k=\pm k_{EP}$, showing that the rank of the matrix is smaller than its physical dimension and that two (or more) eigenvectors coalesce. Hence, the presence and the position of the exceptional points do not change when going from the static Hamiltonian to the Floquet Hamiltonian. Looking at the imaginary part of the spectrum (central row in Fig.~\ref{fig:quasi-energy-band}), we see that the EPs now separate the $\mathcal{PT}$-preserved region from the $\mathcal{PT}$-broken region, with an unstable region for $k<|k^\pm_\mathrm{EP}|$ when pumping to the flat band while there is an unstable region for some momenta larger than $|k^\pm_\mathrm{EP}|$ when pumping to the dispersive band. For this reason, we decide to pump to the flat band for $|k|>|k^\pm_\mathrm{EP}|$ and to the dispersive band for $|k|<|k^\pm_\mathrm{EP}|$.
That way, we recover a situation similar to the static Hamiltonian, where $\mathcal{PT}$-symmetry is broken only exactly at the EPs while the spectrum is otherwise entirely real. \\
Note that, since the resonant pumping follows the bands profile the pumping frequency (and hence the period) becomes momentum-dependent. This deforms the Floquet-BZ from a rectangle $[-\pi,\pi]\times[0,\frac{2\pi}{T}]$ to a new surface with width depending on the energy gap between bands $[-\pi,\pi]\times[0,\frac{2\pi}{T(k)}]$.

\section{Bloch space oscillations}

Using the same convention on the normalization of the biorthogonal basis, we can calculate the probability of the wave function to be measured in the $n$-th eigenspace as the expectation value of the $n$-th eigenspace projector $\Pi_n = \ket{\psi_n^\mathrm{R}}\bra{\psi_n^\mathrm{L}}$. With the diagonal metric tensor $G$ the probabilities are calculated as
\begin{equation}
    p_i(t) = \frac{|c_i(t)|^2}{\sum_n|c_n(t)|^2}, \hspace{30pt} \mathrm{with} \hspace{5pt} c_i(t) = \bra{\psi_i^\mathrm{L}}\psi(t)\rangle.
    \label{eq.probNEW}
\end{equation}
We use this definition of probability to study Rabi oscillations in momentum space in our setup. Everything is done analogously to the traditional Rabi oscillations, but replacing the two levels with three bands and the Hermitian probability with its non-Hermitian counterpart. Since we want to see pumping between the first and the third band we initialize the system in the first band, in real space this corresponds to the state:
\begin{equation}
    \ket{\psi(0)} = \sum_{j=-N/2}^{N/2-1} e^{\ii j \mathrm{k}} \ket{j}\otimes \ket{\psi_{1;\mathrm{k}}^\mathrm{R}},
\end{equation}
with $\ket{\psi(0)}$ being the Fourier transform of the lowest energy eigenstate at momentum $k$. After letting the state $\ket{\psi(0)}$ evolve with the Schr\"odinger equation with the real-space Hamiltonian, we expect the state to remain at the same momentum (since the potential is translationally invariant) and oscillate between the first and the third bands. To verify our expectations, we can use Eq.~\ref{eq.probNEW} to calculate the expectation values as a function of momentum for the three bands. To do so, for each time step, we perform an inverse Fourier transform of the wave function, and, for each allowed momentum, we calculate the biorthogonal probability as in Eq.~\ref{eq.probNEW}. The coefficients $c_i(t)$ are simply obtained by projecting the wave function with fixed momentum onto the unperturbed left eigenstates $\bra{\psi_{i;\mathrm{k}}^\mathrm{L}}$.
\begin{figure}[ht]
    \centering
    \includegraphics[width=\linewidth]{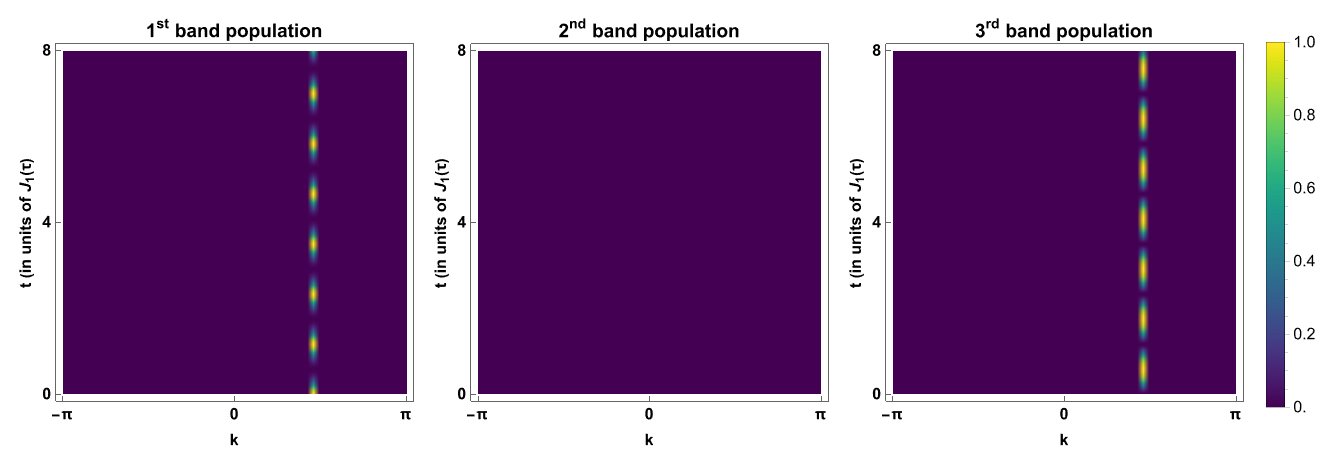}
    \caption{Bi-orthogonal probability of the wave function onto the three bands at $k = \frac{14\pi}{30}$. Each panel gives the bi-orthogonal probability for a different band as a function of both time and momentum. As we see, the probability periodically oscillates between first and third band while the second band is unpopulated. Furthermore, the potential preserves the momentum and the wave function has, at every time, the same momentum of the initial state.}
    \label{fig:prob}
\end{figure}
As we see in Fig.~\ref{fig:prob}, all the other momenta and the second band are completely excluded from the dynamic (provided that we are in resonance or close to it). For this reason, Fig.~\ref{fig_3} can be simplified without plotting all the momenta and all the bands.

\section{Robustness against detuning}
\label{App.detuning}
In this section, we want to compare the pumping efficiency of Hermitian and Non-Hermitian Rabi oscillations. \\
Rabi pumping is typically used in two-level systems to induce oscillations between the $\ket{+}$ and the $\ket{-}$ state, and replacing a single atom with a one dimensional lattice does not introduce any qualitatively new phenomena. However, in the discussed non-Hermitian three band model, the Bloch momentum serves as a tunable parameter that can be used to study the influence of EPs on the dynamics of the system. Eventually, the $\mathcal{PT}$ symmetry ensures stable dynamics with the total power oscillating around an average value~\cite{Bian2020}, as opposite to the Hermitian evolution where the total power is constant. \\
We can also investigate the robustness of the process against detuning of the driving frequency. As we know from theory, a complete population transfer from the initial state to the target state is only possible if the pumping frequency is in resonance with the level spacing. When some detuning is added, both the Rabi frequency and the maximum amplitude are renormalized. \\
In the Hermitian Rabi oscillations we expect the generalized Rabi frequency to be $\Omega_\mathrm{R}'=\sqrt{\Omega_\mathrm{R}^2+\delta^2}$ and the new transferred population to be at most $\Omega_\mathrm{R}^2/(\Omega_\mathrm{R}^2+\delta^2)$~\cite{Foot_2005}. Since the starting hypothesis is that the potential is perturbative and the resulting Rabi frequency is smaller than the typical energy scale, the amplitude of the oscillations is quickly suppressed by driving detuning. \\ 
To generalize these results to our non-Hermitian model we start with the Schr\"odinger equation 
\begin{align}
        \ii\dot{c}_{n';k}=\frac{1}{2\ii}\sum_{n}\braket{\psi^\mathrm{L}_{n';k}|V(k)|\psi^\mathrm{R}_{n;k}}\left(\ee^{-\ii(\Delta\epsilon-\Omega_\mathrm{P})t}-\ee^{-\ii(\Delta\epsilon+\Omega_\mathrm{P})t}\right),
\end{align}
and substitute $\Omega_\mathrm{P}$ with $\Omega_\mathrm{P}+\delta$, with $\delta$ being the detuning. If we assume that the detuning is much smaller than the energy gap $\Delta\epsilon$, we can still use the rotating wave approximation to remove all the terms with high frequency. Eventually, Eq.~\ref{set_equ_Rabi} becomes
\begin{align}
    \ii\dot{c}_{\mathrm{in};k}(t)&=c_{\mathrm{t};k}(t)\frac{1}{2\ii}e^{\ii \delta t}\braket{\psi^\mathrm{L}_{\mathrm{in};k}|V(k)|\psi^\mathrm{R}_{\mathrm{t};k}}\\
    \ii\dot{c}_{\mathrm{t};k}(t)&=-c_{\mathrm{in};k}(t)\frac{1}{2\ii}e^{-\ii\delta t}\braket{\psi^\mathrm{L}_{\mathrm{t};k}|V(k)|\psi^\mathrm{R}_{\mathrm{in};k}}.
    \label{set_equ_Rabi_det}
\end{align}
Similarly to the pumping in resonance, the solution of the differential equation with $c_\mathrm{in,k}(0)=1$ and $c_\mathrm{t;k}(0)=0$ is
\begin{align}
    c_{\mathrm{in};k}(t) = e^{\ii\delta t/2} \left( \cos \left(\frac{\sqrt{\delta^2+\Omega_\mathrm{R}^2}}{2}t  \right) - \ii \frac{\delta}{\sqrt{\delta^2+\Omega_\mathrm{R}^2}} \sin \left( \frac{\sqrt{\delta^2+\Omega_\mathrm{R}^2}}{2}t \right) \right) \\
    c_{\mathrm{t};k}(t) = -e^{-\ii\delta t/2 } \frac{\bra{\psi_{\mathrm{t};k}^L}V(k)\ket{\psi_{\mathrm{in};k}^R}}{\sqrt{\delta^2+\Omega_\mathrm{R}^2}}\sin \left( \frac{\sqrt{\delta^2+\Omega_\mathrm{R}^2}}{2}t\right),
\end{align}
As in the previous section, we calculate the bi-orthogonal transition probabilities by choosing $g_n = 1$ and using the simple equation $p_i(t) = |c_i(t)|^2/\sum_j|c_j(t)|^2$, obtaining eventually
\begin{align}
    p_{\mathrm{in};k}(t) = \dfrac{\cos^2 \dfrac{\Omega_\mathrm{R}'}{2}t + \dfrac{\delta^2}{\Omega_\mathrm{R}'^2}\sin^2\dfrac{\Omega_\mathrm{R}'}{2}t}{\cos^2 \dfrac{\Omega_\mathrm{R}'}{2}t + \dfrac{\delta^2+|V_{\mathrm{in;t}}|^2}{\Omega_\mathrm{R}'^2}\sin^2\dfrac{\Omega_\mathrm{R}'}{2}t} \\
    p_{\mathrm{t};k}(t) = \dfrac{ \dfrac{|V_{\mathrm{in;t}}|^2}{\Omega_\mathrm{R}'^2}\sin^2\dfrac{\Omega_\mathrm{R}'}{2}t}{\cos^2 \dfrac{\Omega_\mathrm{R}'}{2}t + \dfrac{\delta^2+|V_{\mathrm{in;t}}|^2}{\Omega_\mathrm{R}'^2}\sin^2\dfrac{\Omega_\mathrm{R}'}{2}t},
    \label{eq.coeffs}
\end{align}
with $\Omega_\mathrm{R}'=\sqrt{\Omega_\mathrm{R}^2+\delta^2}$ and $V_{\mathrm{in;t}}=\bra{\psi_{\mathrm{t};k}^\mathrm{L}}V(k)\ket{\psi_{\mathrm{in};k}^\mathrm{R}}$. Notice that in the Hermitian case $|V_{\mathrm{in;t}}|=\Omega_\mathrm{R}$ as we do not distinguish between left and right eigenvectors and the denominator is 1, giving the expected results. Furthermore, for $\delta=0$ the Hermitian probabilities simply reduce to the usual sine and cosine functions while in the non-Hermitian case they become deformed. The degree of deformation is strictly related to the degree of self-orthogonality through the term $V_{\mathrm{in;t}}$. It is also easy to see that the higher the Rabi frequency the more robust this process is against detuning such that the maximum value of $p_{\mathrm{t};k}(t)$ decreases much slower as a function of $\delta$. 
\begin{figure}[ht]
    \centering
    \includegraphics[width=0.5\linewidth]{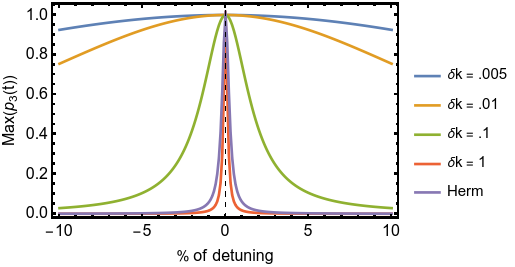}
    \caption{Efficiency of the Rabi pumping as a function of the detuning. The efficiency of the process is defined as the maximum value of population transferred from the initial state to the target state, and it's calculated as the maximum over time of $p_{\mathrm{t};k}(t)$ from Eq.~\ref{eq.coeffs}. The detuning is expressed as a percentage of the pumping frequency $\Omega_\mathrm{P}=\epsilon_3-\epsilon_1$ to give standardized results. The efficiency of the Rabi pumping in the Hermitian case is also displayed for comparison.}
    \label{fig:detuning}
\end{figure}
As we can see in Fig.~\ref{fig:detuning}, the robustness of the Rabi pumping strongly depends on the distance from the EP $\delta k$. When very close to the EP, $\Omega_\mathrm{R}\to\infty$ and the process is very robust, while far from the EP the results are comparable to the Hermitian counterpart, as any trace of self-orthogonality vanishes. 


\end{document}